\documentclass[11pt,a4paper,nofootinbib]{article} 

\usepackage{jheppub}
\usepackage{enumerate}
\usepackage{epsfig}
\usepackage{amsmath} 
\usepackage{capt-of}
\usepackage{wasysym} 
\usepackage{mathrsfs} 
\usepackage{graphicx} 
\usepackage{amsfonts} 
\usepackage{amsbsy} 
\usepackage{amscd}
\usepackage{rotating}
\usepackage{slashed, latexsym, verbatim, cancel} 
\usepackage[utf8]{inputenc}
 \usepackage{multirow}
\usepackage{color}
\definecolor{myred}{rgb}{0.6,0,0} 
\definecolor{myblue}{rgb}{0,0.2,0.4}
\definecolor{mygreen}{rgb}{0,0.9,0.1}
\definecolor{hc}{rgb}{.9,0.1,0.7}
\definecolor{hcout}{rgb}{.9,0.7,0.9}
\definecolor{Orange}{rgb}{1.,0.65,0.}

\makeatletter

\newcommand{\fmslash}[2][0mu]{%
  \mathchoice
    {\fmsl@sh\displaystyle{#1}{#2}}%
    {\fmsl@sh\textstyle{#1}{#2}}%
    {\fmsl@sh\scriptstyle{#1}{#2}}%
    {\fmsl@sh\scriptscriptstyle{#1}{#2}}}
\newcommand{\fmsl@sh}[3]{%
  \m@th\ooalign{$\hfil#1\mkern#2/\hfil$\crcr$#1#3$}}
\makeatother

\newcommand{\lsim}{{\;\raise0.3ex\hbox{$<$\kern-0.75em\raise-1.1ex\hbox{$\sim$}}\;}}
\newcommand{\gsim}{{\;\raise0.3ex\hbox{$>$\kern-0.75em\raise-1.1ex\hbox{$\sim$}}\;}}

\usepackage{array}
\newcolumntype{C}[1]{>{\centering\arraybackslash$}p{#1}<{$}}

\usepackage{bm} 

\usepackage{cleveref}
\newcommand{\be}{\begin{equation}}
\newcommand{\ee}{\end{equation}}
\newcommand{\bes}{\begin{equation*}}
\newcommand{\ees}{\end{equation*}}
\newcommand{\bea}{\begin{eqnarray}}
\newcommand{\eea}{\end{eqnarray}}
\newcommand{\beas}{\begin{eqnarray*}}
\newcommand{\eeas}{\end{eqnarray*}}

\newcommand{\wt}{\widetilde}
\def\ra{\rightarrow}

\title{Reconstructing heavy Higgs boson masses in Type X Two-Higgs Doublet Model
with a light pseudoscalar}
\author[a]{Eung Jin Chun,}
\author[b]{Siddharth Dwivedi,}
\author[b]{Tanmoy Mondal,} 
\author[b]{Biswarup Mukhopadhyaya,} 
\author[b]{Santosh Kumar Rai}
\affiliation[a]{Korea Institute for Advanced Study, Seoul 02455, Korea}
\affiliation[b]{Regional Centre for Accelerator-based Particle Physics,
Harish-Chandra Research Institute, HBNI,
Chhatnag Road, Jhunsi, Allahabad - 211 019, India} 
\emailAdd{ejchun@kias.re.kr}
\emailAdd{siddharthdwivedi@hri.res.in}
\emailAdd{tanmoymondal@hri.res.in}
\emailAdd{biswarup@hri.res.in}
\emailAdd{skrai@hri.res.in}

\abstract{
We analyze the prospects of reconstructing the mass of a heavy charged Higgs boson 
in the context of a Type X two-Higgs doublet model where a light pseudoscalar
$A$ in the mass range $40-60$ GeV is phenomenologically allowed, and is in fact
favoured if one wants to explain the muon anomalous magnetic moment. The 
associated production of charged Higgs with the pseudoscalar $A$ and subsequent
decay of the charged Higgs into a $W$ and $A$, is found to be our 
relevant channel. The branching ratio for $H^+ \to W^+ A$ with $M_{H^+} \sim 200$ GeV, 
is close to 50\%. The hadronic decay of the $W$ boson, coupled with the leptonic decays 
of $A$ into a tau and muon pair, help in identifying the charged Higgs. The neutral heavy
Higgs, being degenerate with the charged Higgs for most of the allowed
parameter space of the model, also contributes to similar final states. 
Thus both of these particles are reconstructed within a band of about 10 GeV.}

\preprint{HRI-RECAPP-2018-006\\$\textrm{}$\hfill \today}


\begin{document}
\maketitle


\section{Introduction}
In describing new physics an extended scalar sector can be of relevance in several contexts 
including supersymmetry, CP-violation and dark matter. Of the possible scenarios,
two-Higgs doublet models (2HDMs) stand out as minimalistic but phenomenologically
rich options, whose signatures can be tested at colliders.
There are four broad categories of 2HDMs which respect natural flavor
conservation at the tree level, due to the presence of some discrete
symmetry in the Lagrangian. These are usually named Type I, Type II, 
Type X (or lepton specific) and Type Y (or flipped) \cite{Gunion:1989we,Djouadi:2005gj,Branco:2011iw}. 
This paper focuses on identifying the collider signatures of the heavy Higgs
bosons in Type X 2HDM, which has a viable region of parameter space
that explains the muon $g-2$ discrepancy \cite{Brown:2001mga, Bennett:2006fi}. 
This region allows for a sufficiently light ($40-60$ GeV) pseudoscalar, 
coupled with a high value of tan $\beta$, that can give enhanced (positive)
two-loop contribution to the anomalous muon magnetic moment \cite{Cheung:2001hz,
Cheung:2003pw,Jegerlehner:2009ry,Broggio:2014mna,Chun:2015hsa}. 
Such low values of $M_A$, the pseudocalar mass, are consistent
with all experimental limits \cite{Kanemura:2014dea,Broggio:2014mna,Chun:2015hsa}.


In this scenario, one scalar doublet has Yukawa
couplings with quarks only, while the other one couples to leptons
alone. This results in the ``hadrophobic'' nature of the couplings
of the heavy Higgs bosons and the pseudoscalar, allowing a light
pseudoscalar to escape detection at LHC. It has been demonstrated
\cite{Broggio:2014mna,Chun:2015hsa,Chun:2016hzs} that the 
neutral pseudoscalar $A$ in type X 2HDM can be as
light at 40-60 GeV or even lighter in certain regions in the
parameter space respecting all the constraints coming from
collider data, muon $g-2$, flavor constraints, electroweak
precision data and theoretical constraints from vacuum stability
and perturbativity. There have been several studies exploring
signatures of the scalar sector of the type X 2HDM at LHC
and $e^+ e^-$ colliders \cite{Su:2009fz,Kanemura:2011kx,Kanemura:2014bqa,Kanemura:2014dea,
Chun:2015hsa}. In a recent work, the issue of reconstructing
such a light pseudocalar was studied \cite{Chun:2017yob}, 
utilizing the decay mode of the pseudoscalar into a muon pair, enabling 
reconstruction of the sharp invariant mass peak.

For large $\tan\beta$, the light pseudoscalar with mass around 
50 GeV has a $\tau^+ \tau^-$ branching ratio close to unity, and a
$\mu^+ \mu^-$ branching ratio of the order of $0.35 \%$.
We consider the channel $p~ p \to H^{\pm} A$, where the charged Higgs decays via 
$H^{\pm} \to W^{\pm} A $ and then the pseudoscalar $A$'s decay to a tau or 
muon pair, {\it i.e.} $A \to \mu^+ \mu^- $ and $A \to \tau^+ \tau^- $. 
The invariant mass reconstruction from the muon pair will clearly be able to identify 
the pseudoscalar with a sharp resonance. 
We show how one can reconstruct the charged Higgs ($H^{\pm}$) and the
heavier neutral scalar ($H$), making use of the $A$-reconstruction strategy 
delineated in \cite{Chun:2017yob}.

In Section 2 we recapitulate the Type X 2HDM and point out how
the parameter space of the model gets constrained by perturbativity 
and vacuum stability, muon $g-2$ and precision observables. 
Section 3 includes the LHC analysis of our signal, detailing the mass
reconstruction strategy and the kinematic distributions used to suppress
the SM background contributions. Section 4 includes a discussion of the
numerical results for different benchmark points used in our analysis.  
We summarize and conclude in Section 5.

\section{The Type X 2HDM}
The Type X 2HDM with two scalar doublets  $\Phi_{1,2}$  has the following
Yukawa Lagrangian:
\begin{equation}\label{eq:yukawa}
{\cal L}_Y= -Y^u\bar{ Q_L} \wt \Phi_2 u_R + Y^d  \bar{ Q_L} \Phi_2 d_R+Y^e\bar{ l_L} \Phi_1 e_R + h.c.,
\end{equation}
where  $\wt \Phi_2=i\sigma_2\Phi_2^*$ and family indices have been suppressed. This Yukawa 
structure results from a $\mathbb{Z}_2$ symmetry~\cite{PhysRevD.15.1958} ensuring
invariance under $\Phi_2\ra \Phi_2$ and $\Phi_1\ra-\Phi_1$ together with $e_R\ra -e_R$ 
, other fermions being even under it. Thus  $\Phi_2$ couples only to quarks and
$\Phi_1$ couples exclusively to the leptons.
The most general 2HDM scalar potential is 
\begin{eqnarray}
\nonumber V_{\mathrm{2HDM}} &=& m_{11}^2\Phi_1^{\dagger}\Phi_1 + m_{22}^2\Phi_2^{\dagger}\Phi_2 -\Big[m_{12}^2\Phi_1^{\dagger}\Phi_2 + \mathrm{h.c.}\Big]
+\frac{1}{2}\lambda_1\left(\Phi_1^\dagger\Phi_1\right)^2+\frac{1}{2}\lambda_2\left(\Phi_2^\dagger\Phi_2\right)^2 \\
\nonumber && +\lambda_3\left(\Phi_1^\dagger\Phi_1\right)\left(\Phi_2^\dagger\Phi_2\right)+\lambda_4\left(\Phi_1^\dagger\Phi_2\right)\left(\Phi_2^\dagger\Phi_1\right)
+\Big\{ \frac{1}{2}\lambda_5\left(\Phi_1^\dagger\Phi_2\right)^2+\Big[\lambda_6\left(\Phi_1^\dagger\Phi_1\right) \\
&& +\lambda_7\left(\Phi_2^\dagger\Phi_2\right)\Big]\left(\Phi_1^\dagger\Phi_2\right) + \rm{h.c.}\Big\},
\label{eq:2hdmgen}
\end{eqnarray}
where all the couplings are assumed to be real to ensure CP-conservation. 
The $\mathbb{Z}_2$ symmetry implies $\lambda_6=\lambda_7=0$. However, we 
allow for soft $\mathbb{Z}_2$ breaking in the potential with a non 
vanishing $m_{12}^2$ term to keep the quartic coupling $\lambda_1$ 
below perturbativity limit \cite{ Gunion:1989we,Gunion:2002zf}.   
Parameterizing the doublets as 
\begin{align}
\Phi_i =   \begin{pmatrix} H^+_i \\ \cfrac{v_i + h_i + i A_i}{\sqrt 2} \end{pmatrix}, ~~~i = 1,2 \,\, ;\label{2hdm_doublets}
\end{align}
we obtain the five massive physical states $A$ (CP-odd), $h$, $H$, $H^{\pm}$ in terms of
the two diagonalizing angles 
$\alpha$ and $\beta$ such that
\begin{align}
 \begin{pmatrix} H  \\ h \end{pmatrix} =  \begin{pmatrix}  c_{\alpha} && s_{\alpha} \\ -s_{\alpha} &&  c_{\alpha} \end{pmatrix}
  \begin{pmatrix} h_1  \\ h_2 \end{pmatrix}  \label{2hdm_scalar_basis}
 \end{align}
and  $A=-s_\beta \;A_1 + c_\beta \;A_2,\quad H^{\pm}=-s_\beta\; H_1^{\pm} + c_\beta\; H^{\pm}_2$
where $s_\alpha = {\rm sin}~\alpha$, $c_\beta = {\rm cos}~ \beta$ etc and ${\rm tan}~\beta = \cfrac{v_2}{v_1}$ .
The CP-even state $h$ is identified with the SM-like Higgs with mass $M_h \approx 125$ GeV.

The Yukawa Lagrangian of Eq.(\ref{eq:yukawa}), when written in terms of the interactions of 
matter fields with the physical Higgs bosons is given by
\begin{eqnarray}
\nonumber \mathcal L_{\mathrm{Yukawa}}^{\mathrm{Physical}} &=&
-\sum_{f=u,d,\ell} \frac{m_f}{v}\left(\xi_h^f\overline{f}hf +
\xi_H^f\overline{f}Hf - i\xi_A^f\overline{f}\gamma_5Af \right) \\
 &&-\left\{ \frac{\sqrt{2}V_{ud}}
{v}\overline{u}\left(m_{u}\xi_A^{u}P_L+m_{d}\xi_A^{d}P_R\right)H^{+}d  +
\frac{\sqrt{2}m_l}{v}\xi_A^l\overline{v}_LH^{+}l_R + \mathrm{h.c.}\right\},
\label{eq:L2hdm}
\end{eqnarray}
where $v = \sqrt{{v_1}^2 + {v_2}^2} = 246$ GeV and  $u$, $d$, and $l$ refer 
to up-type quarks, down-type quarks and charged leptons, respectively. 
The multiplicative factors $\xi_h^f$, $\xi_H^f$ 
and $\xi_A^f$ are listed in Table \ref{Tab:YukawaFactors}. 
\begin{table}[t]
\begin{center}
\begin{tabular}{|c||c|c|c|c|c|c|c|c|c|}
\hline
& $\xi_h^u$ & $\xi_h^d$ & $\xi_h^\ell$
& $\xi_H^u$ & $\xi_H^d$ & $\xi_H^\ell$
& $\xi_A^u$ & $\xi_A^d$ & $\xi_A^\ell$ \\ \hline
Type X
& $c_\alpha/s_\beta$ & $c_\alpha/s_\beta$ & $-s_\alpha/c_\beta$
& $s_\alpha/s_\beta$ & $s_\alpha/s_\beta$ & $c_\alpha/c_\beta$
& $\cot\beta$ & $-\cot\beta$ & $\tan\beta$ \\
 \hline
\end{tabular}
\end{center}
 \caption{The multiplicative factors of Yukawa interactions in Type X 2HDM}
\label{Tab:YukawaFactors}
\end{table}
  
Three point vertices involving the heavy Higgs and the gauge bosons 
relevant to our analysis are ~\cite{Gunion:1989we,Djouadi:2005gj, Kanemura:2014dea} 
\begin{align}
 HAZ_\mu:\,-\frac{g_Z^{}}{2}\sin(\beta-\alpha)(p+p')_\mu,
~~H^\pm AW^\mp_\mu:\,\frac{g}{2}(p+p')_\mu,
\label{hhV}
\end{align}
where $p$ and $p'$ are the outgoing four-momenta of the first
and the second scalars, respectively, and $g_Z=g/\cos\theta_W$. Note that
the coupling of the pseudoscalar $A$ to gauge boson pairs vanishes
due to CP invariance $i.e. ~ g_{AVV} = 0$. The couplings
of the light CP-even Higgs $h$ and the heavy neutral Higgs $H$
to a pair of gauge bosons have the form
\begin{align}
g_{hVV}=\mathrm{sin}(\beta-\alpha)g_{hVV}^{\mathrm{SM}} \, , && g_{HVV}=\mathrm{cos}(\beta-\alpha)g_{hVV}^{\mathrm{SM}} \label{VVh_VVH c}
\end{align}
where $V$ = $Z,\,W^\pm$.
Thus, when $\beta - \alpha \to \frac{\pi}{2}$ (alignment 
limit), the couplings of the lighter CP-even Higgs $h$ approach that
of the SM Higgs while $g_{HVV} \to 0$. 
From Table \ref{Tab:YukawaFactors} we can see the
hadrophobic nature of A for large $\tan\beta$, with $\xi^{u(d)}_A = {\rm cot}~\beta (-{\rm cot}~\beta)$.
This would result in low yield for the $A$ production via gluon fusion, 
which is the dominant production mode at LHC. 
\subsection{Constraints on the model parameters}
From direct searches at LEP there exists a model-independent
limit on the charged Higgs mass of $M_{H^{\pm}} > 79.3$ GeV \cite{Heister:2002ev}.
From flavor observables, Type X escapes the strong constraint of 
$M_{H^{\pm}} > 580$ GeV from $\bar B \to X_s \gamma$, most common in Type II 2HDM
\cite{Belle:2016ufb}. This is because the couplings of $H^{\pm}$
to quarks in Type X 2HDM are proportional to cot $\beta$. 
However a light pseudoscalar of $M_A < 10$ GeV is still ruled out
from $B_s \to \mu^+ \mu^-$ \cite{Wang:2014sda}.

In view of the muon ($g-2$) result, the region of parameter space of
interest to us prefers a light pseudoscalar with $M_A \lesssim 70$ GeV 
with $ \tan{\beta}\gg 1$.
From considerations of perturbativity and vacuum stability \cite{Broggio:2014mna}, 
charged Higgs mass has an upper bound of $M_{H^{\pm}} \lesssim 200$ GeV
for $M_A \lesssim 100$ GeV in the right sign limit of Yukawa modifiers,
{\it i.e.} $\xi^{\ell}_h > 0 $. 
However, it is unconstrained in the wrong sign limit {\it i.e.} 
for $\xi^{\ell}_h < 0 $. Since we are interested in the region
where the pseudoscalar mass is $40-60$ GeV, we are working in
the wrong sign limit \cite{Chun:2015hsa}. Moreover, it has been 
shown using electroweak precision data \cite{Broggio:2014mna} that in the
alignment limit, for nearly degenerate heavy neutral and charged scalars
($H, H^{\pm}$) all values of $M_A$ are permissible. In addition, the choice
of our benchmarks is guided by the requirement to keep the 
branching ratio of $h \to A A$ within 3-4\% so as to satisfy the
exclusion limits provided by the CMS collaboration \cite{Sirunyan:2018mbx}. 

\section{Mass reconstruction strategy: signal and backgrounds}
As stated earlier, the signal channel considered in the analysis here is the associated production of the
charged Higgs boson with the light pseudoscalar at LHC:
\begin{align}
p~ p \to H^{\pm} A \,\, , 
\end{align}
with another $A$ appearing in the final state through $H^{\pm}$ decay ($H^{\pm} \to W^{\pm} A$). 
The pseudoscalar then decays into a tau or muon pair, {\it i.e.} $A \to \mu^+ \mu^- $ 
or $A \to \tau^+ \tau^- $. Note that the heavy neutral Higgs which is nearly degenerate with the 
charged Higgs boson can also be produced in association with $A$ via a $Z$ mediated process 
\begin{align}
p ~p \to H A \,\, . 
\end{align}
This also contributes to the same final state as $H \to Z A$, and therefore has a substantial 
bearing on the total signal strength when the gauge bosons $W$ and $Z$ appearing in the decay cascades above, decay 
hadronically into a pair of jets ($j$). 
The signal is tagged with a final state containing a pair of muons, at least two light jets and 
at least one tau-tagged jet ($j_\tau$). The invariant mass of the heavy Higgs 
(charged or neutral) is identified with the invariant mass of the system consisting of two 
leading jets (not tau-tagged) in $p_T$ reconstructing the weak gauge bosons, and a pair of 
oppositely charged muons. Since the 
muon pair can come from either the associated $A$ or the one via $H^{\pm} (H)$
decay, we need additional cuts to maximize the contribution of the signal to the invariant mass 
of the $2\mu 2j$ system. Note that the signal peaks for $N_j=2$ and therefore the $W/Z$ boson is reconstructed using the
two leading jets only. 

 \begin{figure}[ht!]
 \begin{center}
   \includegraphics[width=0.65\textwidth]{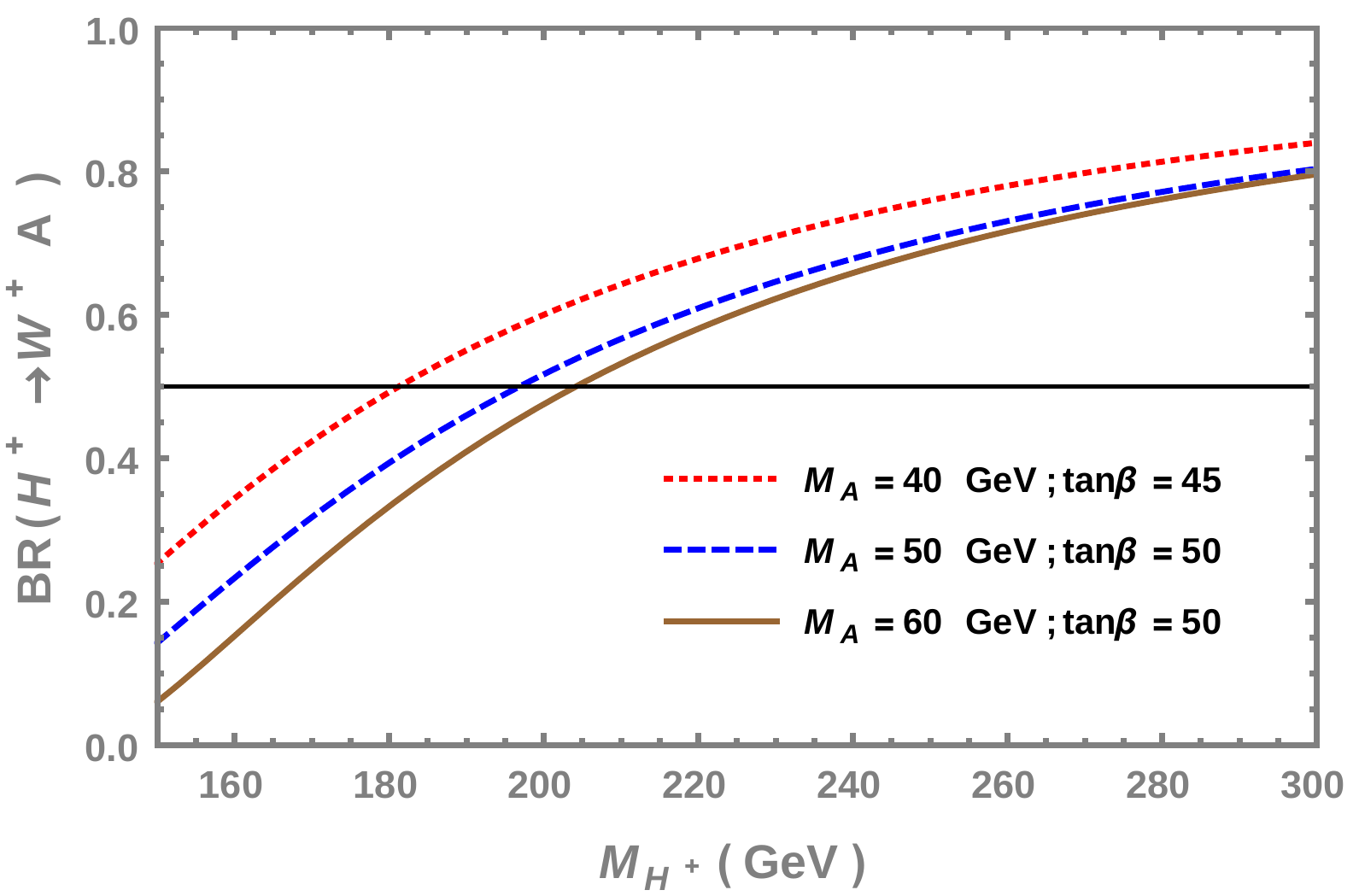}
 \caption{BR($H^{\pm} \to W^{\pm} A$) vs. $M_{H^\pm}$
for  $M_A = 40,~ 50 ~{\rm and}~ 60$ GeV. The horizontal line
represents the 50\% branching ratio.}
  \label{fig:BR_Charged_Higgs}
   \end{center}
  \end{figure}
Our benchmark points include three values of the
pseudoscalar mass, namely, $M_A = 40,~ 50 ~{\rm and}~ 60$ GeV.
For each value of $M_A$ we vary the charged Higgs mass in the
range $150 < M_{H^{\pm}} < 300$ GeV. We tune the value of 
$\tan\beta$ and $\cos(\beta - \alpha)$ to respect the
constraints from $(g-2)_\mu$ and BR($h \to AA$). In the given
range for $M_{H^{\pm}}$, $ H^{\pm} \to W^{\pm} A$ and
$ H^{\pm} \to \tau^+ \nu_\tau$ are the two dominant modes. 
Of these two decay modes, the branching ratio for 
$ H^{\pm} \to W^{\pm} A$ depends on $M_{H^{\pm}},~ M_A$
and $M_W$ but not on tan $\beta$. However,
BR($ H^{\pm} \to \tau^+ \nu_\tau$) is proportional to 
$M_{H^{\pm}}{\rm tan}^2~\beta$ \cite{Chun:2015hsa}. 
Respecting the constraints from lepton universality and muon 
$(g-2)$ \cite{Chun:2016hzs}, higher tan $\beta$ values are 
allowed but increasing tan $\beta$ would cause 
$ H^{\pm} \to \tau^+ \nu_\tau$  to win against the
$ H^{\pm} \to W^{\pm} A$ channel. Keeping this in mind
we tune the value of tan $\beta$ for the different values of
$M_A$ so as to simultaneously satisfy all the constraints
and have BR($H^{\pm} \to W^{\pm} A$) $>$ BR($ H^{\pm} \to \tau^+ \nu_\tau$).
Figure~\ref{fig:BR_Charged_Higgs} shows a variation of
BR($H^{\pm} \to W^{\pm} A$) with $M_{H^\pm}$ for  
 $M_A = 40,~ 50 ~{\rm and}~ 60$ GeV.

\subsection{Backgrounds}
The major contributions to the SM background for our final state $ \mu^+\mu^-\ 2j\ j_\tau $ come from 
(a) $p p \to \mu^+ \mu^- + jets$, (b) $p p \to t \bar t + jets$ and (c) $p p \to V V + jets (V = Z,W, \gamma^*) $.
Of these (a) is the most dominant background having contributions from both the on-shell $Z$ 
as well as the off-shell photon ($\gamma*$) continuum. This is followed by (b) and (c).
All the background events are generated with two additional partons and the events are matched up to 
two jets using MLM matching scheme \cite{Mangano:2006rw,Hoche:2006ph}  using the \emph{shower-kT} algorithm 
with $p_T$ ordered showers. We have used relevant k-factors to account for the QCD radiative corrections to the 
SM subprocesses. 
Apart from the above three subprocesses, $tW + jets$ could also contribute to the 
SM background. However, its contribution was found to be negligible as compared to (a) and (b)
background channels, and is therefore ignored in the analysis.

\subsection{Simulation and event selection}\label{sec:simulation}
  
Signal and background events have been simulated with $\texttt{MadGraph5\_aMC@NLO}$ \cite{Alwall:2011uj,Alwall:2014hca}
fed to \texttt{PYTHIA6} \cite{Sjostrand:2006za} for the subsequent decay, showering and hadronization 
of the parton level events. $\tau$ decays are incorporated via \texttt{TAUOLA} \cite{Jadach:1993hs}
integrated in $\texttt{MadGraph5\_aMC@NLO}$. 
Event generation uses the \texttt{NN23LO1} \cite{Ball:2014uwa} parton distribution 
function and the default dynamic renormalisation and factorisation scales \cite{mad:scale} in $\texttt{MadGraph5\_aMC@NLO}$. 
Finally, detector simulation is incorporated in \texttt{Delphes3} \cite{deFavereau:2013fsa}
using the  anti-kT algorithm \cite{Cacciari:2008gp} for jet reconstruction with $R = 0.4$. 
In \texttt{Delphes3}, the $\tau$-tagging efficiency and mistagging efficiencies of
the light jets as $\tau$-jets are chosen to be the ``Medium tag point" as quoted in 
\cite{ATL-PHYS-PUB-2015-045}. This entails the tagging efficiency of 1-prong (3-prong)
$\tau$ decay to be 70\% (60\%) and the corresponding mistagging rate is 1\% (2\%).  

\begin{figure}[t!]
 \begin{center}
 \includegraphics[width=5.1cm,angle=-90]{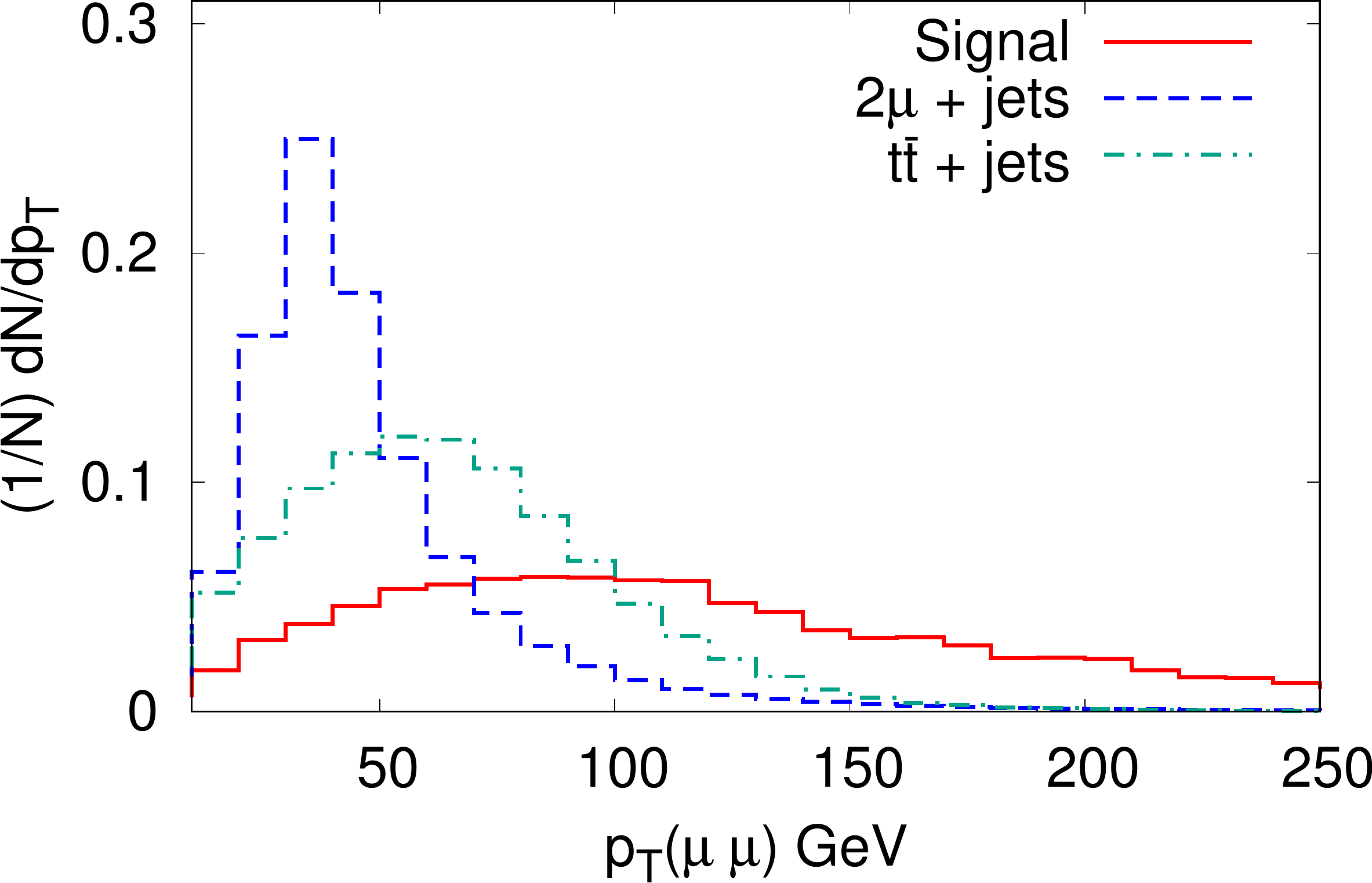}\hspace{-0.2cm}
 \includegraphics[width=5.2cm,angle=-90]{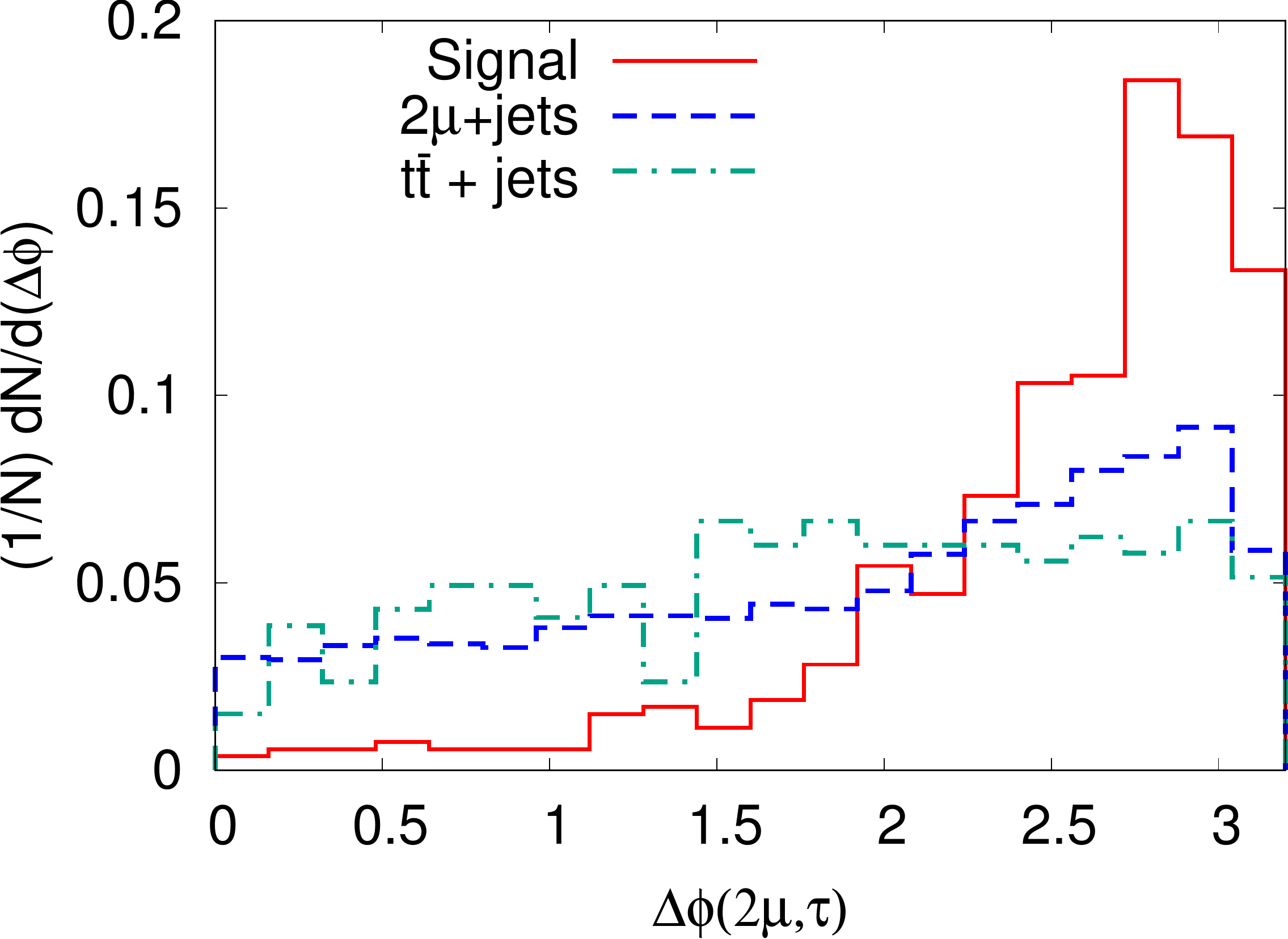}
 \caption{In the left panel we show the $p_T \, (\mu\mu)$ distribution for the signal and 
  background.  In the right panel we show the 
  azimuthal angle separation between the muon pair and the highest $p_T \,\, \tau$-tagged 
  jet.}
 \label{fig:pt-mumu}
 \end{center}
\end{figure} 
We use the following selection cuts to select our signal and reduce the accompanying backgrounds:
\begin{itemize}
  \item {\bf Preselection Cuts (a) : } We require the final state to have two 
  oppositely charged muons with $p_T > 10$ GeV accompanied with two light
  jets and at least one tau-tagged jet of $p_T > 20$ GeV.
  
  \item {\bf Preselection Cuts (b) : } We also demand a $b$-veto on the final state. This helps 
  to suppress the $t \bar t + jets$ and $tW+jets$ background.
  \item The invariant mass of the di-muon system ($M_{\mu\mu}$) satisfies $| M_{\mu\mu}~-~ M_A | < 2.5  \textrm{ GeV}$. 
  \item The $p_T$ of the muon pair has a minimum threshold of $p_T(\mu\mu) > 90$ GeV. 
  This is chosen keeping in mind that the muons coming from the $A$ decay, which
  in turn comes from the $H^{\pm}$ or $H$ decay, are expected to be boosted. 
  The transverse momentum distribution of the muon system is depicted in left panel of 
  Figure~\ref{fig:pt-mumu}.
  The signal events are generated with $M_A = 50$ GeV and $M_{H^\pm/H} = 210$ GeV. 
  It is evident from the Figure~that a cut  of 90 GeV on $p_T(\mu\mu)$ will suppress the background considerably. 
  \item Finally we also impose a minimum azimuthal separation between
  the muon pair and the hardest tau-tagged jet, $i.e.$ $\Delta \phi_{2\mu, j_\tau} > 1.6$.
  This is because the muon pair and the tau-tagged jet are expected to arise from the
  decays of $H^{\pm}$ and the associated $A$ respectively. Thus they are expected to have a large
  azimuthal separation since $H^{\pm}$ and $A$ are expected to be almost back to back
  and therefore well separated. This is depicted in right panel of Figure~\ref{fig:pt-mumu}. 
  It is evident that a cut on $\Delta \phi_{2\mu, j_\tau}$ will reduce substantial 
  amount of the background. 
\end{itemize}
 Note that the leading dijet system in our analysis is also expected to satisfy an invariant mass window
  of $|M_{j_1 j_2} - 85.0| < 20.0 $ GeV about the $W$ or $Z$ resonance, which helps us in reconstructing the 
  heavy Higgs mass.

\section{Results and Discussion}
\begin{table}
\renewcommand{\arraystretch}{1.2}
\centering
\begin{tabular}{|c||c|c||c|c||c|}
\hline
\multirow{2}{*}{Cuts}         & \multicolumn{2}{c|}{Signal} & \multicolumn{2}{c|}{Background}     & Significance\\ \cline{2-5} 
                                    & $H^\pm A$  & $H A$    & $\mu^+\mu^-$+jets & $t\bar{t}$+jets &\\ \hline
Preselection Cuts (a)               & 179        & 79       &  38610            & 25424    & 1.0 \\ \hline
Preselection Cuts (b)               & 173        & 72       &  37755            & 10125     & 1.1\\ \hline
$|M_{\mu\mu}-M_A| < 2.5 $ GeV       & 151        & 63       &  9228             & 2444     & 2.0\\ \hline
$p_T (\mu\mu) > 90$ GeV             & 108        & 44       &  2351             & 605      & 2.8 \\ \hline
$\Delta\Phi(\mu\mu,j_\tau) > 1.6$   & 98         & 40       &  1742              & 354     & 3.0 \\ \hline
\end{tabular}
\caption{Cut flow table displaying effectiveness of different cuts used to enhance signal to background ratio. 
  Signal events are generated with $M_{H^\pm} = M_{H} = 210$ GeV and $M_A = 50$ GeV. All the events are estimated with 
  integrated luminosity of $3000 fb^{-1}$ data.}
\label{tab:cut-flow}
\end{table}

In the previous sections we discussed the analysis framework and simulation cuts
which can be utilized to improve the signal to background ratio. To
quantify the efficacy of different cuts, we consider a benchmark point with 
$M_A = 50$ GeV and $M_{H^\pm} = M_{H} =  210$ GeV and step-by-step cut flow is 
presented in Table~\ref{tab:cut-flow}. The events are estimated with an integrated
luminosity of 3000 $fb^{-1}$. Production cross-section for a 210 GeV charged and neutral
Higgs along with a 50 GeV pseudoscalar is 120 fb and 60 fb respectively. 
  \begin{figure}[b!]
 \begin{center}
 \includegraphics[width=0.85\textwidth]{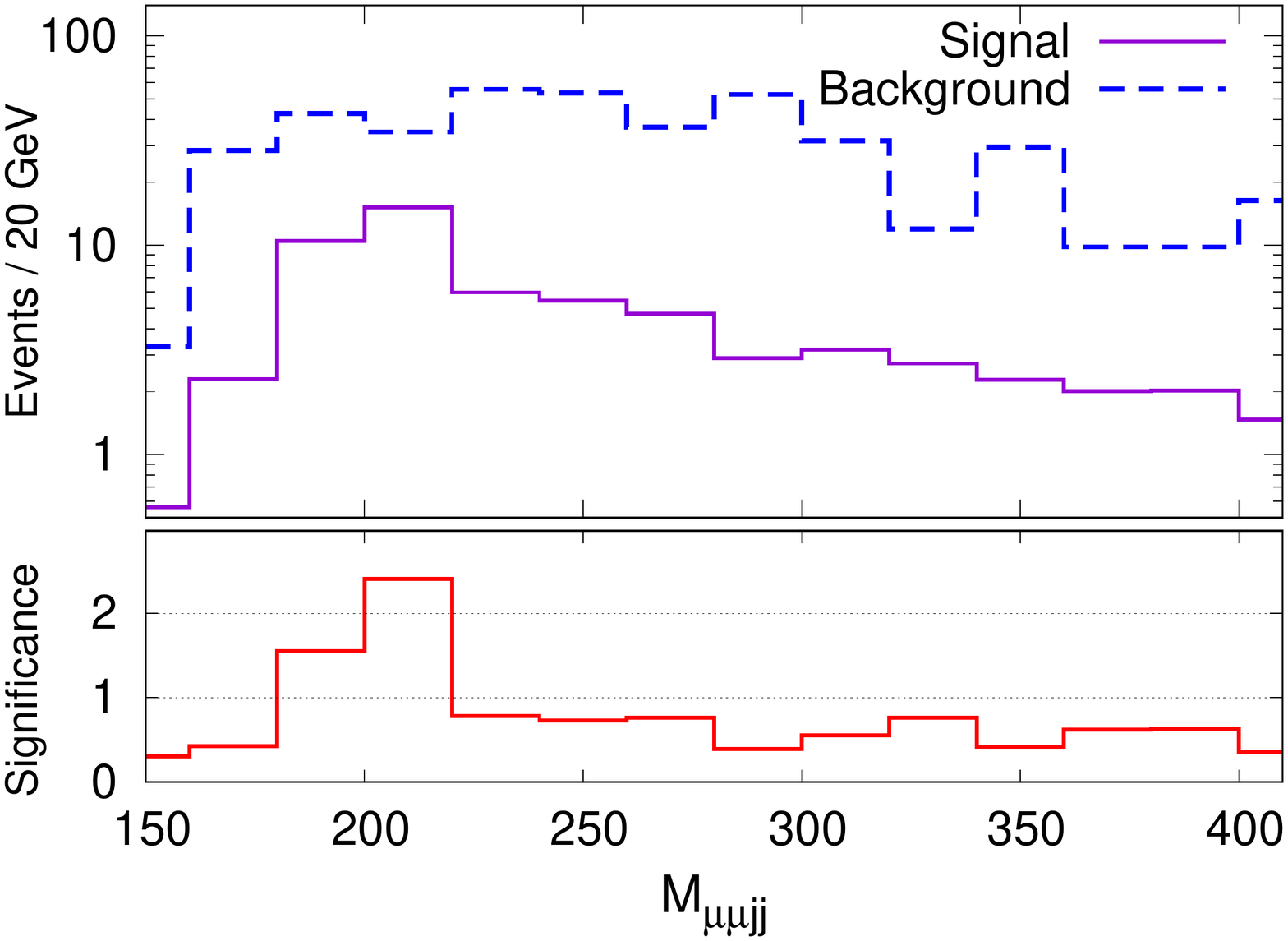}
 \caption{Invariant mass distribution of $\mu^+ \,\mu^- \,j\,j$ system for signal and background 
 events. 
Signal event are generated for heavy scalar mass of 210 GeV with $M_A = 50 $GeV.  
The bottom panel shows the binwise significance of the signal comparing the total events to the estimated background 
events in each invariant mass bin.
 }
 \label{fig:reconstruct}
 \end{center}
\end{figure}
With these cuts we have analyze the signal $(S)$ and 
background events $(B)$, and present the corresponding statistical significance $(\mathcal{S})$ 
at each step in the right most column. We estimate the significance using the expression:
\begin{equation}
\mathcal{S} = \sqrt{2\left[(S+B)~\textrm{ln}\left(1+\frac{S}{B}\right)-S\right]}.    %
\end{equation}
It is clear from the event counts in Table~\ref{tab:cut-flow} that a search for charged Higgs in the 
mass range of 200 GeV in Type X 2HDM will be quite challenging. A prior knowledge of the 
pseudoscalar mass, which in our case is motivated by $(g-2)_\mu$ data, enables us to devise 
specific selection criteria that helps us achieve only a reasonable significance ($\sim 3\sigma$) 
for its observation. We now aim to reconstruct the mass of the charged Higgs with enough confidence in 
that particular mass window.  
To do this, we have plotted the invariant mass distribution of $\mu\,\mu\,j\,j$ system for the signal and background events in 
Figure~\ref{fig:reconstruct}. Note that for signal events, we have merged events 
coming from both the charged Higgs and heavy neutral Higgs production channels. 
The  'Background' events represent the sum of $t\bar t+\,jets$ and $2\mu+\,jets$ processes. 
The signal events are generated for heavy scalar mass of 210 GeV with $M_A = 50 $ GeV. 
In the bottom panel of Figure~\ref{fig:reconstruct} we show the local significance calculated 
for each bin of the invariant mass using the total events to the 
estimated background events in each such bin. Although the actual event shapes of the signal and background in 
the invariant mass distribution when combined may not give a clear indication of a significant 
resonant behavior, the local significance does indicate a clear peak at 210 GeV (mass of heavy scalars) at a 
robust $\simeq 2.2\sigma$. 
\begin{figure}[ht!]
 \begin{center}
 \includegraphics[width=0.65\textwidth]{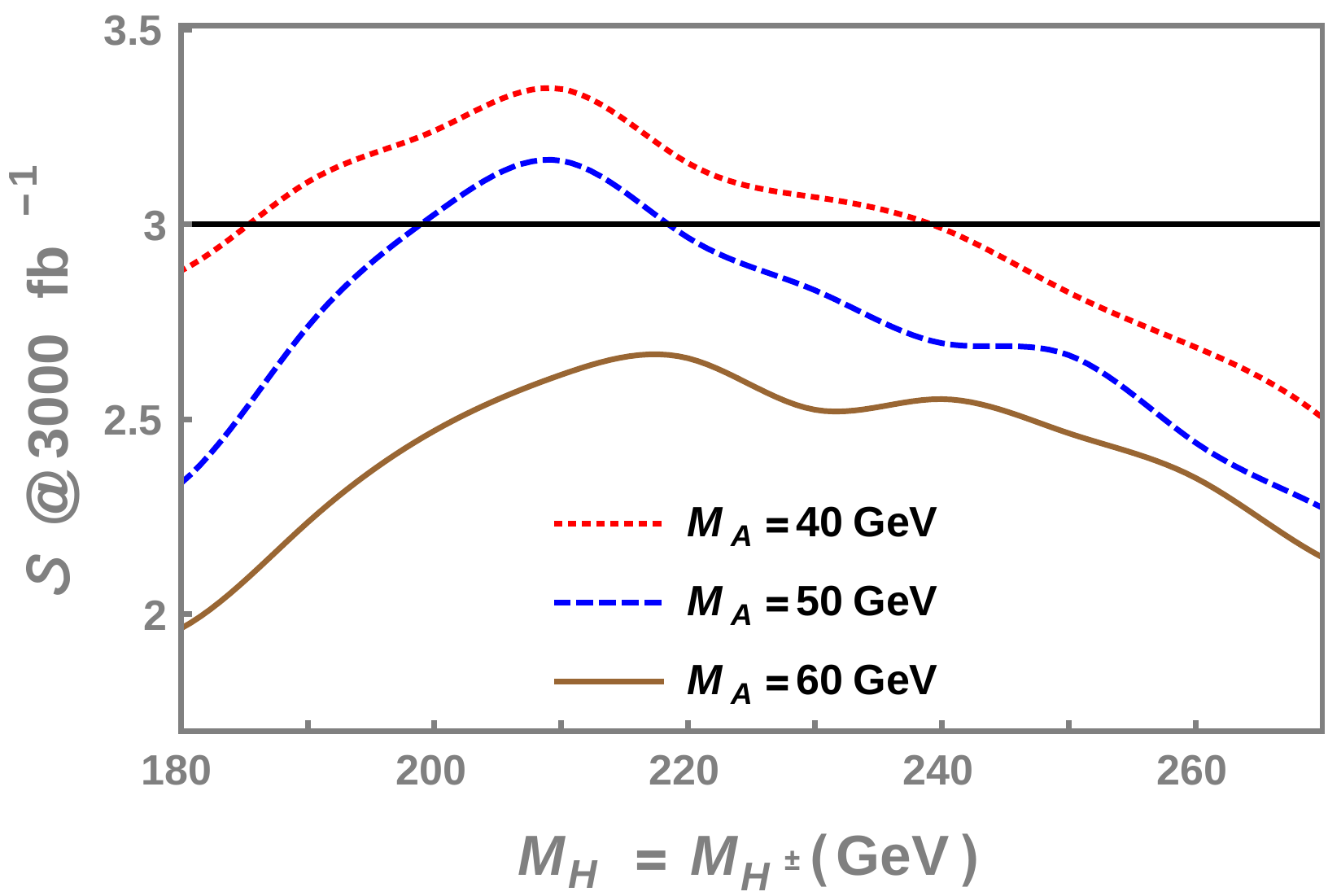}
 \caption{Significance vs. $M_{H^\pm}$
 for  $M_A = 40,~ 50 ~{\rm and}~ 60$ GeV. The horizontal line
 represents the 3$\sigma$ limit.}
 \label{fig:Significance}
 \end{center}
\end{figure}

Now to explore a more general parameter space in $M_{H^\pm}-M_A$ plane,  we vary the charged 
Higgs mass from 180 GeV to 270 GeV and estimate their signal significance. To arrest the effect of 
the pseudoscalar mass, we have analyzed the signal for three different values of the 
pseudoscalar mass, {\it viz.} 40 GeV, 50 GeV and 60 GeV for every choice of the charged Higgs 
mass. Using the same cuts as described in Table~\ref{tab:cut-flow} we have estimated 
the signal significance at 14 TeV LHC with an integrated luminosity of 3000 $fb^{-1}$. 
The variation of the statistical significance as a function of $M_{H^\pm}$ for different values of $M_A$ 
is shown in Figure~\ref{fig:Significance}. 
As the charged Higgs mass increases, the cross section decreases leading to lower significance and 
the same observation is true for pseudoscalar mass. Although the production cross-section is higher 
for light charged Higgs, the branching of $H^\pm$ to $W^\pm A$ is 
low (see Figure~\ref{fig:BR_Charged_Higgs}) which effectively decreases the overall 
significance. We find that the best significance is achieved for moderate values of charged Higgs 
mass i.e. around 200 - 220 GeV where the production cross-section is not very low 
while the branching ratio ($H^\pm \to W^\pm A$) is reasonably large. Thus it would definitely benefit the 
charged Higgs search in Type X 2HDM if LHC were to accumulate more data than the 3000 $fb^{-1}$ .

\section{Summary and Conclusion}

We have succesfully demonstrated the reconstructability of the charged and heavy neutral Higgs
within the Type-X 2HDM scenario, under the assumption of degeneracy of $M_{H^\pm}$ and $M_H$. 
In considering the channel $p~ p \to H^{\pm} A$, and subsequent decays of 
$H^{\pm} \to W^{\pm} A , \,\, W^{\pm} \to j j $, with
$A \to \mu^+ \mu^- ~ {\rm or} ~ \tau^+ \tau^-$, we have taken advantage of the favourable branching
ratio of $ H^{\pm} \to W^{\pm} A $ for heavier $H^\pm$. We have investigated the 
kinematic cuts that can help in suppressing the dominant backgrounds to our final state. To this
end, the sharp invariant mass peak of the di-muon system around the pseudoscalar mass and
a tight $p_T$ threshold on the muon pair is found to be effective in containing the $2\mu + jets$
and $t \bar t + jets $ backgrounds. In addition, invariant mass window on the dijet system 
around the electroweak gauge boson masses also helps in the reconstruction of the heavy 
charged Higgs mass. The contribution coming from the heavy neutral Higgs production to our 
signal yield is found to be relevant as it happens to be nearly half of that of the charged Higgs 
production for the given selection criteria. It is seen that with the increase
in the mass of the pseudoscalar from 40 to 60 GeV, the statistical significance diminishes and a heavy charged
Higgs in the mass range of 200-220 GeV with $M_A = 40$ GeV has the maximum discovery potential.
The analysis projects a significance of $\gtrsim 3\sigma$ for 3000 $fb^{-1}$ for the above benchmark
scenario, which can further improve with a possible luminosity upgrade in the 14 TeV run.
For example, 5000 $fb^{-1}$ may hike the significance close to about 4 $\sigma$.

\section{Acknowledgements}\label{sec:Acknowledgements}
We thank Nabarun Chakrabarty for helpful discussions.
This work was partially supported by funding available from
the Department of Atomic Energy, Government of India, for the Regional Centre for
Accelerator-based Particle Physics (RECAPP), Harish-Chandra Research Institute. 
We also acknowledge the use of the cluster computing setup available at 
RECAPP and at the High Performance Computing facility of HRI.
EJC thanks RECAPP for hospitality where part of the project was finalized.
BM thanks Korea Institute for Advanced Study for hospitality while this project 
was initiated. 

\bibliographystyle{apsrev}
\bibliography{2hdm_charged_Higgs}

\begin{thebibliography}{33}
\expandafter\ifx\csname natexlab\endcsname\relax\def\natexlab#1{#1}\fi
\expandafter\ifx\csname bibnamefont\endcsname\relax
  \def\bibnamefont#1{#1}\fi
\expandafter\ifx\csname bibfnamefont\endcsname\relax
  \def\bibfnamefont#1{#1}\fi
\expandafter\ifx\csname citenamefont\endcsname\relax
  \def\citenamefont#1{#1}\fi
\expandafter\ifx\csname url\endcsname\relax
  \def\url#1{\texttt{#1}}\fi
\expandafter\ifx\csname urlprefix\endcsname\relax\def\urlprefix{URL }\fi
\providecommand{\bibinfo}[2]{#2}
\providecommand{\eprint}[2][]{\url{#2}}

\bibitem[{\citenamefont{Gunion et~al.}(2000)\citenamefont{Gunion, Haber, Kane,
  and Dawson}}]{Gunion:1989we}
\bibinfo{author}{\bibfnamefont{J.~F.} \bibnamefont{Gunion}},
  \bibinfo{author}{\bibfnamefont{H.~E.} \bibnamefont{Haber}},
  \bibinfo{author}{\bibfnamefont{G.~L.} \bibnamefont{Kane}}, \bibnamefont{and}
  \bibinfo{author}{\bibfnamefont{S.}~\bibnamefont{Dawson}},
  \bibinfo{journal}{Front. Phys.} \textbf{\bibinfo{volume}{80}},
  \bibinfo{pages}{1} (\bibinfo{year}{2000}).

\bibitem[{\citenamefont{Djouadi}(2008)}]{Djouadi:2005gj}
\bibinfo{author}{\bibfnamefont{A.}~\bibnamefont{Djouadi}},
  \bibinfo{journal}{Phys. Rept.} \textbf{\bibinfo{volume}{459}},
  \bibinfo{pages}{1} (\bibinfo{year}{2008}), \eprint{hep-ph/0503173}.

\bibitem[{\citenamefont{Branco et~al.}(2012)\citenamefont{Branco, Ferreira,
  Lavoura, Rebelo, Sher, and Silva}}]{Branco:2011iw}
\bibinfo{author}{\bibfnamefont{G.~C.} \bibnamefont{Branco}},
  \bibinfo{author}{\bibfnamefont{P.~M.} \bibnamefont{Ferreira}},
  \bibinfo{author}{\bibfnamefont{L.}~\bibnamefont{Lavoura}},
  \bibinfo{author}{\bibfnamefont{M.~N.} \bibnamefont{Rebelo}},
  \bibinfo{author}{\bibfnamefont{M.}~\bibnamefont{Sher}}, \bibnamefont{and}
  \bibinfo{author}{\bibfnamefont{J.~P.} \bibnamefont{Silva}},
  \bibinfo{journal}{Phys. Rept.} \textbf{\bibinfo{volume}{516}},
  \bibinfo{pages}{1} (\bibinfo{year}{2012}), \eprint{1106.0034}.

\bibitem[{\citenamefont{Brown et~al.}(2001)}]{Brown:2001mga}
\bibinfo{author}{\bibfnamefont{H.~N.} \bibnamefont{Brown}} \bibnamefont{et~al.}
  (\bibinfo{collaboration}{Muon g-2}), \bibinfo{journal}{Phys. Rev. Lett.}
  \textbf{\bibinfo{volume}{86}}, \bibinfo{pages}{2227} (\bibinfo{year}{2001}),
  \eprint{hep-ex/0102017}.

\bibitem[{\citenamefont{Bennett et~al.}(2006)}]{Bennett:2006fi}
\bibinfo{author}{\bibfnamefont{G.~W.} \bibnamefont{Bennett}}
  \bibnamefont{et~al.} (\bibinfo{collaboration}{Muon g-2}),
  \bibinfo{journal}{Phys. Rev.} \textbf{\bibinfo{volume}{D73}},
  \bibinfo{pages}{072003} (\bibinfo{year}{2006}), \eprint{hep-ex/0602035}.

\bibitem[{\citenamefont{Cheung et~al.}(2001)\citenamefont{Cheung, Chou, and
  Kong}}]{Cheung:2001hz}
\bibinfo{author}{\bibfnamefont{K.-m.} \bibnamefont{Cheung}},
  \bibinfo{author}{\bibfnamefont{C.-H.} \bibnamefont{Chou}}, \bibnamefont{and}
  \bibinfo{author}{\bibfnamefont{O.~C.~W.} \bibnamefont{Kong}},
  \bibinfo{journal}{Phys. Rev.} \textbf{\bibinfo{volume}{D64}},
  \bibinfo{pages}{111301} (\bibinfo{year}{2001}), \eprint{hep-ph/0103183}.

\bibitem[{\citenamefont{Cheung and Kong}(2003)}]{Cheung:2003pw}
\bibinfo{author}{\bibfnamefont{K.}~\bibnamefont{Cheung}} \bibnamefont{and}
  \bibinfo{author}{\bibfnamefont{O.~C.~W.} \bibnamefont{Kong}},
  \bibinfo{journal}{Phys. Rev.} \textbf{\bibinfo{volume}{D68}},
  \bibinfo{pages}{053003} (\bibinfo{year}{2003}), \eprint{hep-ph/0302111}.

\bibitem[{\citenamefont{Jegerlehner and Nyffeler}(2009)}]{Jegerlehner:2009ry}
\bibinfo{author}{\bibfnamefont{F.}~\bibnamefont{Jegerlehner}} \bibnamefont{and}
  \bibinfo{author}{\bibfnamefont{A.}~\bibnamefont{Nyffeler}},
  \bibinfo{journal}{Phys. Rept.} \textbf{\bibinfo{volume}{477}},
  \bibinfo{pages}{1} (\bibinfo{year}{2009}), \eprint{0902.3360}.

\bibitem[{\citenamefont{Broggio et~al.}(2014)\citenamefont{Broggio, Chun,
  Passera, Patel, and Vempati}}]{Broggio:2014mna}
\bibinfo{author}{\bibfnamefont{A.}~\bibnamefont{Broggio}},
  \bibinfo{author}{\bibfnamefont{E.~J.} \bibnamefont{Chun}},
  \bibinfo{author}{\bibfnamefont{M.}~\bibnamefont{Passera}},
  \bibinfo{author}{\bibfnamefont{K.~M.} \bibnamefont{Patel}}, \bibnamefont{and}
  \bibinfo{author}{\bibfnamefont{S.~K.} \bibnamefont{Vempati}},
  \bibinfo{journal}{JHEP} \textbf{\bibinfo{volume}{11}}, \bibinfo{pages}{058}
  (\bibinfo{year}{2014}), \eprint{1409.3199}.

\bibitem[{\citenamefont{Chun et~al.}(2015)\citenamefont{Chun, Kang, Takeuchi,
  and Tsai}}]{Chun:2015hsa}
\bibinfo{author}{\bibfnamefont{E.~J.} \bibnamefont{Chun}},
  \bibinfo{author}{\bibfnamefont{Z.}~\bibnamefont{Kang}},
  \bibinfo{author}{\bibfnamefont{M.}~\bibnamefont{Takeuchi}}, \bibnamefont{and}
  \bibinfo{author}{\bibfnamefont{Y.-L.~S.} \bibnamefont{Tsai}},
  \bibinfo{journal}{JHEP} \textbf{\bibinfo{volume}{11}}, \bibinfo{pages}{099}
  (\bibinfo{year}{2015}), \eprint{1507.08067}.

\bibitem[{\citenamefont{Kanemura
  et~al.}(2014{\natexlab{a}})\citenamefont{Kanemura, Yokoya, and
  Zheng}}]{Kanemura:2014dea}
\bibinfo{author}{\bibfnamefont{S.}~\bibnamefont{Kanemura}},
  \bibinfo{author}{\bibfnamefont{H.}~\bibnamefont{Yokoya}}, \bibnamefont{and}
  \bibinfo{author}{\bibfnamefont{Y.-J.} \bibnamefont{Zheng}},
  \bibinfo{journal}{Nucl. Phys.} \textbf{\bibinfo{volume}{B886}},
  \bibinfo{pages}{524} (\bibinfo{year}{2014}{\natexlab{a}}),
  \eprint{1404.5835}.

\bibitem[{\citenamefont{Chun and Kim}(2016)}]{Chun:2016hzs}
\bibinfo{author}{\bibfnamefont{E.~J.} \bibnamefont{Chun}} \bibnamefont{and}
  \bibinfo{author}{\bibfnamefont{J.}~\bibnamefont{Kim}},
  \bibinfo{journal}{JHEP} \textbf{\bibinfo{volume}{07}}, \bibinfo{pages}{110}
  (\bibinfo{year}{2016}), \eprint{1605.06298}.

\bibitem[{\citenamefont{Su and Thomas}(2009)}]{Su:2009fz}
\bibinfo{author}{\bibfnamefont{S.}~\bibnamefont{Su}} \bibnamefont{and}
  \bibinfo{author}{\bibfnamefont{B.}~\bibnamefont{Thomas}},
  \bibinfo{journal}{Phys. Rev.} \textbf{\bibinfo{volume}{D79}},
  \bibinfo{pages}{095014} (\bibinfo{year}{2009}), \eprint{0903.0667}.

\bibitem[{\citenamefont{Kanemura et~al.}(2012)\citenamefont{Kanemura, Tsumura,
  and Yokoya}}]{Kanemura:2011kx}
\bibinfo{author}{\bibfnamefont{S.}~\bibnamefont{Kanemura}},
  \bibinfo{author}{\bibfnamefont{K.}~\bibnamefont{Tsumura}}, \bibnamefont{and}
  \bibinfo{author}{\bibfnamefont{H.}~\bibnamefont{Yokoya}},
  \bibinfo{journal}{Phys. Rev.} \textbf{\bibinfo{volume}{D85}},
  \bibinfo{pages}{095001} (\bibinfo{year}{2012}), \eprint{1111.6089}.

\bibitem[{\citenamefont{Kanemura
  et~al.}(2014{\natexlab{b}})\citenamefont{Kanemura, Tsumura, Yagyu, and
  Yokoya}}]{Kanemura:2014bqa}
\bibinfo{author}{\bibfnamefont{S.}~\bibnamefont{Kanemura}},
  \bibinfo{author}{\bibfnamefont{K.}~\bibnamefont{Tsumura}},
  \bibinfo{author}{\bibfnamefont{K.}~\bibnamefont{Yagyu}}, \bibnamefont{and}
  \bibinfo{author}{\bibfnamefont{H.}~\bibnamefont{Yokoya}},
  \bibinfo{journal}{Phys. Rev.} \textbf{\bibinfo{volume}{D90}},
  \bibinfo{pages}{075001} (\bibinfo{year}{2014}{\natexlab{b}}),
  \eprint{1406.3294}.

\bibitem[{\citenamefont{Chun et~al.}(2017)\citenamefont{Chun, Dwivedi, Mondal,
  and Mukhopadhyaya}}]{Chun:2017yob}
\bibinfo{author}{\bibfnamefont{E.~J.} \bibnamefont{Chun}},
  \bibinfo{author}{\bibfnamefont{S.}~\bibnamefont{Dwivedi}},
  \bibinfo{author}{\bibfnamefont{T.}~\bibnamefont{Mondal}}, \bibnamefont{and}
  \bibinfo{author}{\bibfnamefont{B.}~\bibnamefont{Mukhopadhyaya}},
  \bibinfo{journal}{Phys. Lett.} \textbf{\bibinfo{volume}{B774}},
  \bibinfo{pages}{20} (\bibinfo{year}{2017}), \eprint{1707.07928}.

\bibitem[{\citenamefont{Glashow and Weinberg}(1977)}]{PhysRevD.15.1958}
\bibinfo{author}{\bibfnamefont{S.~L.} \bibnamefont{Glashow}} \bibnamefont{and}
  \bibinfo{author}{\bibfnamefont{S.}~\bibnamefont{Weinberg}},
  \bibinfo{journal}{Phys. Rev. D} \textbf{\bibinfo{volume}{15}},
  \bibinfo{pages}{1958} (\bibinfo{year}{1977}),
  \urlprefix\url{https://link.aps.org/doi/10.1103/PhysRevD.15.1958}.

\bibitem[{\citenamefont{Gunion and Haber}(2003)}]{Gunion:2002zf}
\bibinfo{author}{\bibfnamefont{J.~F.} \bibnamefont{Gunion}} \bibnamefont{and}
  \bibinfo{author}{\bibfnamefont{H.~E.} \bibnamefont{Haber}},
  \bibinfo{journal}{Phys. Rev.} \textbf{\bibinfo{volume}{D67}},
  \bibinfo{pages}{075019} (\bibinfo{year}{2003}), \eprint{hep-ph/0207010}.

\bibitem[{\citenamefont{Heister et~al.}(2002)}]{Heister:2002ev}
\bibinfo{author}{\bibfnamefont{A.}~\bibnamefont{Heister}} \bibnamefont{et~al.}
  (\bibinfo{collaboration}{ALEPH}), \bibinfo{journal}{Phys. Lett.}
  \textbf{\bibinfo{volume}{B543}}, \bibinfo{pages}{1} (\bibinfo{year}{2002}),
  \eprint{hep-ex/0207054}.

\bibitem[{\citenamefont{Abdesselam et~al.}(2016)}]{Belle:2016ufb}
\bibinfo{author}{\bibfnamefont{A.}~\bibnamefont{Abdesselam}}
  \bibnamefont{et~al.} (\bibinfo{collaboration}{Belle}), in
  \emph{\bibinfo{booktitle}{{Proceedings, 38th International Conference on High
  Energy Physics (ICHEP 2016): Chicago, IL, USA, August 3-10, 2016}}}
  (\bibinfo{year}{2016}), \eprint{1608.02344},
  \urlprefix\url{https://inspirehep.net/record/1479946/files/arXiv:1608.02344.pdf}.

\bibitem[{\citenamefont{Wang and Han}(2015)}]{Wang:2014sda}
\bibinfo{author}{\bibfnamefont{L.}~\bibnamefont{Wang}} \bibnamefont{and}
  \bibinfo{author}{\bibfnamefont{X.-F.} \bibnamefont{Han}},
  \bibinfo{journal}{JHEP} \textbf{\bibinfo{volume}{05}}, \bibinfo{pages}{039}
  (\bibinfo{year}{2015}), \eprint{1412.4874}.

\bibitem[{\citenamefont{Sirunyan et~al.}(2018)}]{Sirunyan:2018mbx}
\bibinfo{author}{\bibfnamefont{A.~M.} \bibnamefont{Sirunyan}}
  \bibnamefont{et~al.} (\bibinfo{collaboration}{CMS Collaboration}),
  \bibinfo{type}{Tech. Rep.} \bibinfo{number}{CMS-HIG-17-029-003},
  \bibinfo{institution}{CERN}, \bibinfo{address}{Geneva}
  (\bibinfo{year}{2018}), \urlprefix\url{http://cds.cern.ch/record/2317389}.

\bibitem[{\citenamefont{Mangano et~al.}(2007)\citenamefont{Mangano, Moretti,
  Piccinini, and Treccani}}]{Mangano:2006rw}
\bibinfo{author}{\bibfnamefont{M.~L.} \bibnamefont{Mangano}},
  \bibinfo{author}{\bibfnamefont{M.}~\bibnamefont{Moretti}},
  \bibinfo{author}{\bibfnamefont{F.}~\bibnamefont{Piccinini}},
  \bibnamefont{and} \bibinfo{author}{\bibfnamefont{M.}~\bibnamefont{Treccani}},
  \bibinfo{journal}{JHEP} \textbf{\bibinfo{volume}{01}}, \bibinfo{pages}{013}
  (\bibinfo{year}{2007}), \eprint{hep-ph/0611129}.

\bibitem[{\citenamefont{Hoeche et~al.}(2005)\citenamefont{Hoeche, Krauss,
  Lavesson, Lonnblad, Mangano, Schalicke, and Schumann}}]{Hoche:2006ph}
\bibinfo{author}{\bibfnamefont{S.}~\bibnamefont{Hoeche}},
  \bibinfo{author}{\bibfnamefont{F.}~\bibnamefont{Krauss}},
  \bibinfo{author}{\bibfnamefont{N.}~\bibnamefont{Lavesson}},
  \bibinfo{author}{\bibfnamefont{L.}~\bibnamefont{Lonnblad}},
  \bibinfo{author}{\bibfnamefont{M.}~\bibnamefont{Mangano}},
  \bibinfo{author}{\bibfnamefont{A.}~\bibnamefont{Schalicke}},
  \bibnamefont{and} \bibinfo{author}{\bibfnamefont{S.}~\bibnamefont{Schumann}},
  in \emph{\bibinfo{booktitle}{{HERA and the LHC: A Workshop on the
  implications of HERA for LHC physics: Proceedings Part A}}}
  (\bibinfo{year}{2005}), pp. \bibinfo{pages}{288--289},
  \eprint{hep-ph/0602031},
  \urlprefix\url{http://inspirehep.net/record/709818/files/arXiv:hep-ph_0602031.pdf}.

\bibitem[{\citenamefont{Alwall et~al.}(2011)\citenamefont{Alwall, Herquet,
  Maltoni, Mattelaer, and Stelzer}}]{Alwall:2011uj}
\bibinfo{author}{\bibfnamefont{J.}~\bibnamefont{Alwall}},
  \bibinfo{author}{\bibfnamefont{M.}~\bibnamefont{Herquet}},
  \bibinfo{author}{\bibfnamefont{F.}~\bibnamefont{Maltoni}},
  \bibinfo{author}{\bibfnamefont{O.}~\bibnamefont{Mattelaer}},
  \bibnamefont{and} \bibinfo{author}{\bibfnamefont{T.}~\bibnamefont{Stelzer}},
  \bibinfo{journal}{JHEP} \textbf{\bibinfo{volume}{06}}, \bibinfo{pages}{128}
  (\bibinfo{year}{2011}), \eprint{1106.0522}.

\bibitem[{\citenamefont{Alwall et~al.}(2014)\citenamefont{Alwall, Frederix,
  Frixione, Hirschi, Maltoni, Mattelaer, Shao, Stelzer, Torrielli, and
  Zaro}}]{Alwall:2014hca}
\bibinfo{author}{\bibfnamefont{J.}~\bibnamefont{Alwall}},
  \bibinfo{author}{\bibfnamefont{R.}~\bibnamefont{Frederix}},
  \bibinfo{author}{\bibfnamefont{S.}~\bibnamefont{Frixione}},
  \bibinfo{author}{\bibfnamefont{V.}~\bibnamefont{Hirschi}},
  \bibinfo{author}{\bibfnamefont{F.}~\bibnamefont{Maltoni}},
  \bibinfo{author}{\bibfnamefont{O.}~\bibnamefont{Mattelaer}},
  \bibinfo{author}{\bibfnamefont{H.~S.} \bibnamefont{Shao}},
  \bibinfo{author}{\bibfnamefont{T.}~\bibnamefont{Stelzer}},
  \bibinfo{author}{\bibfnamefont{P.}~\bibnamefont{Torrielli}},
  \bibnamefont{and} \bibinfo{author}{\bibfnamefont{M.}~\bibnamefont{Zaro}},
  \bibinfo{journal}{JHEP} \textbf{\bibinfo{volume}{07}}, \bibinfo{pages}{079}
  (\bibinfo{year}{2014}), \eprint{1405.0301}.

\bibitem[{\citenamefont{Sjostrand et~al.}(2006)\citenamefont{Sjostrand, Mrenna,
  and Skands}}]{Sjostrand:2006za}
\bibinfo{author}{\bibfnamefont{T.}~\bibnamefont{Sjostrand}},
  \bibinfo{author}{\bibfnamefont{S.}~\bibnamefont{Mrenna}}, \bibnamefont{and}
  \bibinfo{author}{\bibfnamefont{P.~Z.} \bibnamefont{Skands}},
  \bibinfo{journal}{JHEP} \textbf{\bibinfo{volume}{05}}, \bibinfo{pages}{026}
  (\bibinfo{year}{2006}), \eprint{hep-ph/0603175}.

\bibitem[{\citenamefont{Jadach et~al.}(1993)\citenamefont{Jadach, Was, Decker,
  and Kuhn}}]{Jadach:1993hs}
\bibinfo{author}{\bibfnamefont{S.}~\bibnamefont{Jadach}},
  \bibinfo{author}{\bibfnamefont{Z.}~\bibnamefont{Was}},
  \bibinfo{author}{\bibfnamefont{R.}~\bibnamefont{Decker}}, \bibnamefont{and}
  \bibinfo{author}{\bibfnamefont{J.~H.} \bibnamefont{Kuhn}},
  \bibinfo{journal}{Comput. Phys. Commun.} \textbf{\bibinfo{volume}{76}},
  \bibinfo{pages}{361} (\bibinfo{year}{1993}).

\bibitem[{\citenamefont{Ball et~al.}(2015)}]{Ball:2014uwa}
\bibinfo{author}{\bibfnamefont{R.~D.} \bibnamefont{Ball}} \bibnamefont{et~al.}
  (\bibinfo{collaboration}{NNPDF}), \bibinfo{journal}{JHEP}
  \textbf{\bibinfo{volume}{04}}, \bibinfo{pages}{040} (\bibinfo{year}{2015}),
  \eprint{1410.8849}.

\bibitem[{mad()}]{mad:scale}
\bibinfo{howpublished}{\url{"http://cp3.irmp.ucl.ac.be/projects/madgraph/wiki/FAQ-General-13"}}.

\bibitem[{\citenamefont{de~Favereau et~al.}(2014)\citenamefont{de~Favereau,
  Delaere, Demin, Giammanco, Lemaître, Mertens, and
  Selvaggi}}]{deFavereau:2013fsa}
\bibinfo{author}{\bibfnamefont{J.}~\bibnamefont{de~Favereau}},
  \bibinfo{author}{\bibfnamefont{C.}~\bibnamefont{Delaere}},
  \bibinfo{author}{\bibfnamefont{P.}~\bibnamefont{Demin}},
  \bibinfo{author}{\bibfnamefont{A.}~\bibnamefont{Giammanco}},
  \bibinfo{author}{\bibfnamefont{V.}~\bibnamefont{Lemaître}},
  \bibinfo{author}{\bibfnamefont{A.}~\bibnamefont{Mertens}}, \bibnamefont{and}
  \bibinfo{author}{\bibfnamefont{M.}~\bibnamefont{Selvaggi}}
  (\bibinfo{collaboration}{DELPHES 3}), \bibinfo{journal}{JHEP}
  \textbf{\bibinfo{volume}{02}}, \bibinfo{pages}{057} (\bibinfo{year}{2014}),
  \eprint{1307.6346}.

\bibitem[{\citenamefont{Cacciari et~al.}(2008)\citenamefont{Cacciari, Salam,
  and Soyez}}]{Cacciari:2008gp}
\bibinfo{author}{\bibfnamefont{M.}~\bibnamefont{Cacciari}},
  \bibinfo{author}{\bibfnamefont{G.~P.} \bibnamefont{Salam}}, \bibnamefont{and}
  \bibinfo{author}{\bibfnamefont{G.}~\bibnamefont{Soyez}},
  \bibinfo{journal}{JHEP} \textbf{\bibinfo{volume}{04}}, \bibinfo{pages}{063}
  (\bibinfo{year}{2008}), \eprint{0802.1189}.

\bibitem[{ATL(2015)}]{ATL-PHYS-PUB-2015-045}
\bibinfo{type}{Tech. Rep.} \bibinfo{number}{ATL-PHYS-PUB-2015-045},
  \bibinfo{institution}{CERN}, \bibinfo{address}{Geneva}
  (\bibinfo{year}{2015}), \urlprefix\url{https://cds.cern.ch/record/2064383}.

\end{thebibliography}

\end{document}